\documentclass[journal]{IEEEtran}
\ifCLASSINFOpdf
\else
\fi
\usepackage{amssymb,amsmath,latexsym}
\usepackage{pst-all}
\usepackage[varg]{pxfonts}
\usepackage{mathrsfs}
\usepackage{shuffle}
\usepackage{graphicx}

\newtheorem{Definition}{Definition}
\newtheorem{theorem}{Theorem}
\newtheorem{lemma}[theorem]{Lemma}

\newtheorem{Remark}[theorem]{Remark}

\newtheorem{example}[theorem]{Example}
\newtheorem{proposition}[theorem]{Proposition}

\begin{document}
%
\title{Model-Checking of Linear-Time Properties Based on Possibility Measure}
%
%
%

\author{Yongming Li,
      \  Lijun Li

\IEEEcompsocitemizethanks{\IEEEcompsocthanksitem College of Computer Science, Shaanxi Normal University, Xi'an, 710062, China.
 710062.\protect\\
E-mail: liyongm@snnu.edu.cn
\IEEEcompsocthanksitem This work was partially supported by National Science
Foundation of China (Grant No: 11271237,61228305) and the Higher School Doctoral
Subject Foundation of Ministry of Education of China (Grant No:200807180005).}
\thanks{Manuscript received Feb. 8, 2012; revised Aug. 12, 2012.}}

%
%

\markboth{IEEE Transactions on Fuzzy Systems,~Vol., No., November~2012}%
{Shell \MakeLowercase{\textit{et al.}}: Bare Demo of IEEEtran.cls for Journals}
%



\maketitle

\begin{abstract}
Using possibility measure, we study model-checking of linear-time properties in possibilistic Kripke structures. First, the notion of possibilistic Kripke structures and the related possibility measure are introduced, then model-checking of reachability and repeated reachability linear-time properties in finite possibilistic Kripke structures are studied. Standard safety properties and $\omega-$regular properties in possibilistic Kripke structures are introduced, the verification of regular safety properties and $\omega-$regular properties using finite automata are thoroughly studied. It has been shown that the verification of regular safety properties and $\omega-$regular properties in a finite possibilistic Kripke structure can be transformed into the verification of reachability properties and repeated reachability properties in the product possibilistic Kripke structure introduced in this paper. Several examples are given to illustrate the methods presented in the paper.
\end{abstract}

\begin{IEEEkeywords}
Possibilistic Kripke structure, possibility measure, model checking, linear temporal logic, regular language.
\end{IEEEkeywords}

%
\IEEEpeerreviewmaketitle

\section{Introduction}
%
%

%
%
%
%
\IEEEPARstart{I}{n} the last four decades, computer scientists have systematically developed theories of correctness and safety in different aspects, such as methodologies, techniques and even automatic tools for correctness and safety verification of computer systems; see for examples \cite{BF85,MOD99,Lamport94,Pnueli77}.  Of which, model checking has been established as one of the most effective automated techniques for analyzing correctness of software and hardware designs \cite{CJP08,MOD99,McMillan93}. A model checker checks a finite-state system against a correctness 	 property expressed in a propositional temporal logic such as Linear Temporal Logic $(LTL)$ or Computation Tree Logic $(CTL)$. These logics can express safety ($e.g.$,  No two processes can be in the critical section at the same time) and liveness ($e.g.$, Every job sent to the printer will eventually print) properties \cite{BF85, Kripke64,Lamport94,Lamport77,Pnueli77,MYP94}. Model checking has been effectively applied to reasoning about correctness of hardware, communication protocols, software requirements, etc. Many industrial model checkers have been developed, including SPIN \cite{Holzmann97}, SMV \cite{McMillan93}.	

Whereas model-checking techniques focus on the absolute guarantee of correctness - ``it is impossible that the system fails''- in practice such rigid notions are hard, or even impossible, to guarantee. Instead, systems are subject to various phenomena of an uncertainty nature, such as message incomplete or garbling and the like, and correctness - ``with  99 percent chance the system will not fail'' or `` the system will not fail most often'' - is becoming less absolute. To handle with the systematic verification which has something to do with uncertainties in probability, Hart and Sharir in 1986  \cite{SM86}  investigated the logic of timing sequence in probability propositions and applied probability theory to model checking in which the uncertainty is modeled by probability measure. In 2008, Baier and Katoen \cite{CJP08} systematically introduced the principle and method of model checking based on probability measure and related applications with Markov chain models for probability systems.
	
On the other hand, since Zadeh proposed the theory of fuzzy sets in 1965 (\cite{Zadeh65}), many scholars have been devoting themselves to the research in this theory and its applications. As a branch of the theory of fuzzy sets, possibility measure (\cite{Zadeh78}) (more general, fuzzy measure (\cite{Sugeno74})) is a development of classical measure, which focuses on non-additive cases (c.f. \cite{EPRE10,yager12}) that is different from the probability measure which is additive. Most problems in real situations are complicated and non-additive. As a matter of fact, fuzziness seems to pervade most human perception and thinking processes as noted by Zadeh, especially, modeling human-centered systems, for example, biomedical systems (\cite{lin02}), criminal trial systems, decision making systems(\cite{grabisch00}), linguistic quantifiers (\cite{ying06}). Therefore, it is necessary to do some research work in the theory and applications of model checking on non-deterministic systems of non-additive measure, especially, fuzzy measure. And this paper attempts to initiate an $LTL$ model checking based on possibility measure.

	In this paper, the notion of possibilistic Kripke structure is introduced by combining the system with fuzzy uncertainty,  then a possibility measure is induced by the given possibilistic Kripke structure. Linear-time properties specify the traces that a possibilistic Kripke structure should exhibit. Informally speaking, one could say that a linear-time property specifies the admissible (or desired) behavior of the system under consideration. In the following we provide a formal definition of such properties. This definition is rather elementary, and gives an example of what a linear-time property is. In particular, the possibilities of model checking of reachability and repeat reachability are studied, which can be proceeded by solving certain fuzzy relation equations using the least fixed-point method or by constructing the transitive closure of transition possibility distribution. Safety properties and $\omega$-regular properties in possibilistic Kripke structures are introduced. Some calculation methods related to model checking of the above linear-time properties using possibility measure are discussed. In fact, by introducing the product possibilistic Kripke structure, it is shown that model checking of regular safety properties and $\omega$-regular properties in a possibilistic Kripke structure can be calculated by the possibility of reachability and repeated reachability properties of the product possibilistic Kripke structure.

	The rest of the paper is organized as follows. In Section \ref{sec1:PKandmes}, we provide our main definition of possibilistic Kripke structure, and a possibility measure on its paths is introduced. In Section \ref{sec4}, linear-time properties in a possibilistic Kripke structure are introduced, the possibility measures of reachability and repeated reachability properties are studied. The model-checking of regular safety and $\omega$-regular linear-time properties in possibilistic Kripke structures using finite automata are studied in Section \ref{sec5}.  The illustrative example using the treatment of animal is given in Section 5. The conclusion is given in Section 6.


\section{Possibilistic Kripke structure and possibility measure}\label{sec1:PKandmes}
Transition systems or Kripke structures (\cite{Kripke64}) are key models for model checking. Corresponding to possibilistic model checking, we  introduce the notion of possibilistic Kripke structure as follows.

\subsection{Possibilistic Kripke structure}\label{sec1:PKs}

\begin{Definition}\label{sec1:pKripke}
{\rm A possibilistic Kripke structure is a tuple $M=(S,P,I,AP,L)$, where

(1) $S$ is a countable, nonempty set of states;

(2) $P: S\times S\longrightarrow [0,1]$  is the transition possibility distribution such that for all states $s\in S$, $\vee_{s'\in S}P(s,s')=1$;

(3) $I: S\longrightarrow [0,1]$  is the initial distribution, such that $\vee_{s\in S}I(s)=1$;

(4) $AP$ is a set of atomic propositions;

(5) $L: S\longrightarrow 2^{AP}$ is a labeling function that labels a state $s$ with those atomic propositions in $AP$ that are supposed to hold in $s$, where the power set $2^{AP}$ denotes the set of all subsets of $AP$.}

\end{Definition}

	Furthermore, if $S$ and $AP$ are finite sets, then $M=(S,P,I,AP,L)$ is called a finite possibilistic Kripke structure.

	\begin{Remark}\label{sec1:remark1} {\rm (1) In the definition of possibilistic Kripke structure $M$, the labeling function is classical and has no uncertainty. This requirement is not necessary. In fact, we can require the labeling function $L$ to be a function from the state set $S$ to $[0,1]^{AP}$, where $[0,1]^{AP}$ denotes all fuzzy subsets of $AP$, which contains the fuzzy uncertainty. In this case, we can transform it into a possibilistic Kripke structure $M^{\prime}=(S^{\prime}, AP, I^{\prime}, P^{\prime}, L^{\prime})$ as in Definition \ref{sec1:pKripke} as follows, let $D=Im(L)=\{ L(s)(A) | s\in S, A\in AP\}$, $S^{\prime}=S\times D$, and $I^{\prime}(s,d)=I(s)$ for any $(s,d)\in S\times D$, $P^{\prime}((s,d), (s^{\prime},d^{\prime}))=P(s,s^{\prime})$, $L^{\prime}(s,d)=\{A\in AP | L(s,A)\geq d\}$. Then the property of $M$ can be obtained by that of $M^{\prime}$.

(2) In Definition \ref{sec1:pKripke}, we require the transition possibility distribution and initial distribution are normal, i.e., $\vee_{s'\in S}P(s,s')=1$ and $\vee_{s\in S}I(s)=1$, where we use $\vee X$ or $\wedge X$ to represent the least upper bound (or supremum) or the largest lower bound (or infimum) of the subset $X\subseteq [0,1]$, respectively. These conditions are corresponding to the transition probability distribution and probability initial distribution in probabilistic Kripke structure (\cite{CJP08}), where the supremum operation is replaced by the sum operation. They form the main differences between possibilistic Kripke structure and probabilistic Kripke structure. In fact, in fuzzy uncertainty, the order instead of the additivity is one of the most important factors to be considered.

}
\end{Remark}

	The states $s$ with $I(s)>0$ are considered as the initial states. For state $s$ and $T\subseteq S$, let $P(s,T)$ denote the possibility of moving from $s$ to some state $t\in T$ in a single step, that is,
\begin{equation*}
P(s,T)=\vee_{t\in T}P(s,t).
\end{equation*}

	Paths in possibilistic Kripke structure $M$ are infinite paths in the underlying digraph. They are defined as infinite state sequence $\pi=s_0s_1s_2\cdots\in S^\omega$ such that $P(s_i,s_{i+1})>0$ for all $i\in I$. Let $Paths(M)$ denote the set of all paths in $M$, and $Paths_{fin}(M)$ denotes the set of finite path fragments $s_0s_1\cdots s_n$ where $n\geq 0$ and $P(s_i,s_{i+1})>0$ for $0\leq i\leq n-1$ . Let $Paths(s)$ denote the set of all paths in $M$ that start in state $s$. Similarly $Paths_{fin}(s)$ denotes the set of finite path fragments $s_0s_1\cdots s_n$ such that $s_0=s$ .

	For a state $s$, the set of direct successors  (written as $Post(s)$ ) and direct predecessors  (written $Pre(s)$)  are defined as follows:
\begin{itemize}
\item
$ Post(s)=\{s'\in S|P(s,s')>0\}$;
\item
$ Pre(s)=\{s'\in S|P(s',s)>0\}$;
\item
$ Post^*(s)=\{s'\in S|$ there exists $\pi=s_0s_1\cdots s_k\in Paths_{fin}(s)$ such that $s_0=s,s_k=s'\}$;
\item
$ Pre^*(s)=\{s'\in S|s\in Post^*(s')\}$.
\end{itemize}

For $B\subseteq S$, $Post^*(B)=\bigcup_{s\in B}Post^*(s)$, $Pre^*(B)=\bigcup_{s\in B}Pre^*(s)$.

   \begin{Remark}\label{sec1:remark2}
    {\rm  For a possibilistic Kripke structure $M=(S,P,I,AP,L)$, the transition possibility distribution $P$ is a fuzzy relation on $S$,
 its transitive closure $P^+$, say, is also a transition distribution defined as follows,
 $ P^+(s,t)=\vee\{P(s_0,s_1)\wedge\cdots \wedge P(s_{k-1},s_k)|
  s_0=s,s_k=t, s_1,\cdots,s_{k-1}\in S,k\geq1\}$  for $s,t\in S$. Using $P^+$, we can construct a new possibilistic Kripke structure $M^+$, say, as $M^+=(S,P^+,I,AP,L)$. Furthermore, if $M$ is finite, then for any $s,t$ in $S$, there exists a finite state sequence $s_0s_1\cdots s_k$, which is written as $m(s,t)$ in the following, i.e., $m(s,t)=s_0s_1\cdots s_k$, such that $P^+(s,t)=\wedge^{k-1}_{i=0}P(s_i,s_{i+1})$ , we also use $P(m(s,t))$ to represent  $P^+(s,t)=\wedge^{k-1}_{i=0}P(s_i,s_{i+1})$.}
\end{Remark}

\begin{figure}[ht]
\begin{center}
\includegraphics[scale=0.5]{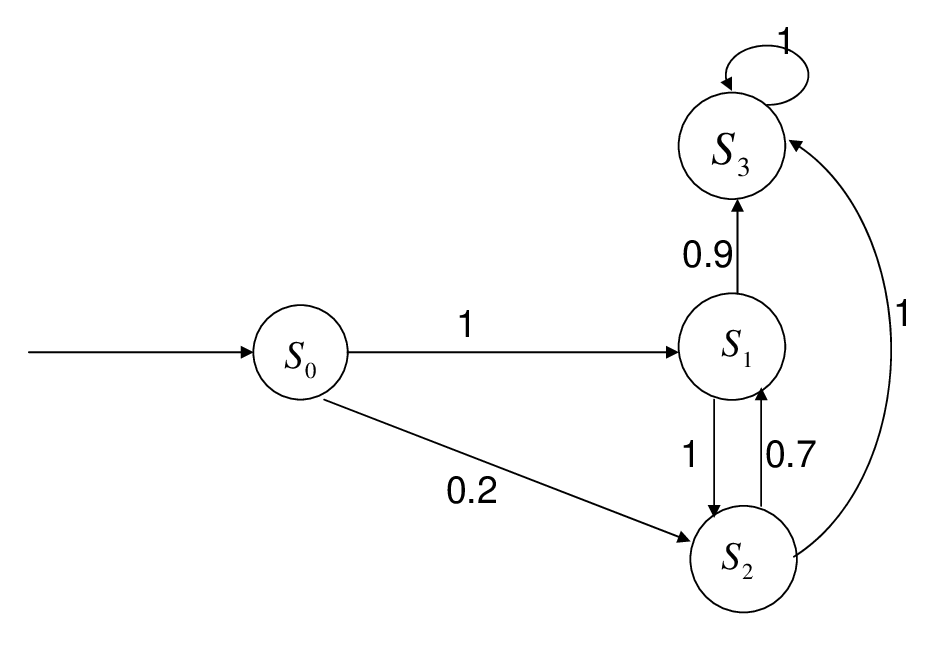}
\center{Fig.1.}A possibilistic Kripke structure $M$ with four states
\vspace{-0.3cm}
\end{center}
\end{figure}

\begin{example}\label{sec1:ep1}
    {\rm  Fig.1 represents a possibilistic Kripke structure $M=(S,P,I,AP,L)$, in which states are represented by ovals and transitions by labeled edges, and $AP=S$.
State names are depicted inside the ovals. Initial states are indicated by having an incoming arrow without source. The state space is $S=\{s_0,s_1,s_2,s_3\}$, the set of initial states consists of only one state $s_0$ such that $I(s_0)=1$. The transition possibility distribution is $P(s_0,s_1)=1$, $P(s_0,s_2)=0.2, P(s_1,s_2)=1, P(s_1,s_3)=0.9, P(s_2,s_1)=0.7 ,P(s_3,s_3)=1, P(s_2,s_3)=1$. The labeling function are $L(s_0)=\{s_0\}$, $L(s_1)=\{s_1\}$, $L(s_2)=\{s_2\}$, $L(s_3)=\{s_3\}$. In the sequel, we often identify the transition possibility distribution $P:S\times S\longrightarrow[0,1]$ with the matrix $(P(s,t))_{s,t\in S}$. Similarly, the initial distribution $I:S\longrightarrow[0,1]$ is often viewed as a vector $(I(s))_{s\in S}$. Using the state order $s_0<s_1<s_2<s_3$,  the matrix $P$ and the vector $I$ are given by
$A=\left(\begin{array}{cccc}0&1&0.2&0\\
0&0&1&0.9\\
0&0.7&0&1\\
0&0&0&1
\end{array}
\right)$ and
$I=\left(\begin{array}{ccc}1\\0\\0\\0
\end{array}
\right)$ .}
\end{example}

     \begin{Remark}
    {\rm A matrix is called a fuzzy matrix if all its elements are taken from the unit interval $[0,1]$. The composition operation
of fuzzy matrixes is similar to ordinary matrix multiplication operation, just let ordinary multiplication and addition operations of real numbers be replaced by minimum and maximum operations of real numbers, which is called $max-min$ composition operation(\cite{Zadeh78}). We use the symbol $\circ$ to represent the $max-min$ composition operation. Then for a possibilistic distribution $P$, its transitive closure is, $P^+=P\vee P^2 \vee \cdots=\vee^\infty_{i=1}P^i$ , where $P^{k+1}=P^k\circ P$ for any positive integer number $k$. When $M$ is a finite possibilistic Kripke structure, the number of states is $n$, say, then $P^+=\vee^n_{i=1}P^i$ (\cite{Li05}).}
\end{Remark}

\subsection{Possibility measure induced by a possibilistic Krikpe structure}

     \begin{Definition}\label{sec1:algebra}
   {\rm  A $(\sigma-)$algebra is a pair $(X,\Omega)$ where $X$ is a nonempty set and $\Omega$ is a set consisting of subsets of $X$
that contains the empty set and is closed under complementation and (countable) unions. Then $(X,\Omega)$ is called a measurable space.}
\end{Definition}

	Given a measurable space $(X,\Omega)$  , recall that a possibility measure (\cite{Zadeh78}) over the algebra $\Omega$ is a mapping $m:\Omega\longrightarrow[0,1]$  satisfying the following conditions,

  (i) $m(\emptyset)=0, m(X)=1$.

(ii) If $A\subseteq B$ in $\Omega$, then $m(A)\leq m(B)$.

(iii) If $A_i\in\Omega,i\in I$, then $Po(\cup_{i\in I}A_i)=\vee_{i\in I}Po(A_i)$.

Sugeno \cite{Sugeno74} calls a mapping $m:\Omega\longrightarrow[0,1]$ a fuzzy measure over $\Omega$ if it satisfies (1) and (ii). Therefore, a possibility measure is also a fuzzy measure.

If $m$ is a possibility measure over the powerset $2^X$, then $m$ is determined by its behaviors on singletons, i.e., if $\Omega=2^X$, then

$m(A)=max_{a\in A}m(\{a\})$ \hfill (1)

\noindent for any subset $A$ of $X$.

\begin{Remark}\label{sec:possibility measure}
{\rm (1) Possibility measure is a class of non-additive measure, it is close related to probability measure. Possibility measure is useful for design in the presence of uncertainties which involve modeling linguistic imprecision or uncertainties with little numercial data to develop a valid probabilistic model. It plays an important role in most human perception and thinking process, especially, modeling human-centered systems, biomedical systems (\cite{lin02}), criminal trial, decision making (\cite{grabisch00}), linguistic quantifiers (\cite{ying06}). As an example, let us see its application in the diagnosing an ill patient with incomplete data.

An ill patient has symptoms of pneumonia, bronchitis, emphysema and cold, but the data of information is incomplete. A doctor gives an estimation by his experience using possibility measure as follows,

$\frac{0.5}{pneumonia}+\frac{0.3}{bronchitis}+\frac{0}{emphysema}+\frac{1}{cold}$,

\noindent which can be extended onto the powerset $\Omega=2^X$ using Eq. (1), where $X=\{pneumonia, bronchitis, emphysema, cold\}$.

(2) Possibility measure is extensional while probability measure is intensional. This makes possibility measures easier to compute than probability measures (\cite{Dra95}). This is also an advantage of using of possibility measures in model checking.}
\end{Remark}

      In the following, we give a possibility measure over a possibilistic Kripke structure $M$.

      \begin{Definition}(\cite{CJP08}) \label{sec:the cylinder}
       {\rm Given a Kripke structure $M$, the cylinder set of $\hat{\pi}=s_0\cdots s_n\in Paths_{fin}(M)$ is defined as,
\begin{equation*}
Cyl(\hat{\pi})=\{\pi\in Paths(M)|\hat{\pi}\in Pref(\pi)\},
\end{equation*}
where $Pref(\pi)=\{\pi'\in Paths_{fin}(M) |\pi'$ is  a finite prefix  of $\pi\}$.}
\end{Definition}

   \begin{Remark} \label{sec:rem:algebra}
  {\rm (1) Assume that $M$ is a possibilistic Kripke structure, then $\Omega=2^{Paths(M)}$  is the algebra generated by
$\{Cyl(\hat{\pi})|\hat{\pi\in Paths_{fin}(M)}\}$ on $Paths(M)$.

     \begin{proof}
       For any $\pi\in Paths(M)$, let $\pi=s_0s_1s_2\cdots$. For all $i\geq0$, write $\pi_i=s_0s_1\cdots s_i$. Obviously,
 $\bigcap_{i=0}^\infty Cyl(\hat{\pi_i})=\{\pi\}$, so $\{\pi\}\in\Omega$ holds for all $\pi\in Paths(M)$. For any $A\subseteq Paths(M),
 A=\bigcup\{\{\pi\}|\pi\in A\}$, so $\Omega=2^{Paths(M)}$.
 \end{proof}

 	(2) If $M$ has at least two elements, then $Paths(M)$ has size continuum, and the $\sigma$-algebra $\Omega_\sigma$ generated by $\{Cyl(\hat{\pi})|$ $\hat{\pi\in Paths_{fin}(M)}\}$ has size at most continuum. In this case, $2^{Paths(M)}$ has size larger than continuum, so the $\sigma$-algebra $\Omega_\sigma$ generated by $\{Cyl(\hat{\pi})|$ $\hat{\pi\in Paths_{fin}(M)}\}$ is not $2^{Paths(M)}$. Furthermore, by the observation of the proof of Remark \ref{sec:rem:algebra}(1), for any $\pi\in Paths(M)$, $\{\pi\}\in \Omega_\sigma$, and the following facts hold,

 (2-1) For any two elements $\hat{\pi_1}$,$\hat{\pi_2}\in Paths_{fin}(M)$,

 \begin{eqnarray*}
 Cyl(\hat{\pi_1})\cap Cyl(\hat{\pi_2})=\left\{
\begin{array}{cc}
Cyl(\hat{\pi_2}),& \hat{\pi_1}\in Pref(\hat{\pi_2});\\
Cyl(\hat{\pi_1}),&\hat{\pi_2}\in Pref(\hat{\pi_1});\\
\emptyset,& otherwise.\\
\end{array}
\right.
 \end{eqnarray*}

 (2-2) If $E=Cyl(s_0\cdots s_k)\in\Omega$, then $E^c=Paths(M)-E$ $=\cup_{i=1}^k\cup_{s\neq s_i}Cyl\{s_0\cdots s_{i-1}s|s\in S\}\in\Omega$.}

\end{Remark}

 For these reasons, we can define a possibility measure on $\Omega=2^{Paths(M)}$.

             \begin{Definition}\label{sec1:Def func}
      {\rm  For a possibilistic Kripke structure $M$, let $Paths(M)=\cup_{s\in S}Paths(s)$. A function
$Po^M: Paths(M)\longrightarrow[0,1]$ is defined as follows: for any $\pi\in Paths(M)$, $\pi=s_0s_1\cdots, Po^M(\pi)=I(s_0)\wedge\bigwedge_{i=1}^\infty P(s_i,s_{i+1})$. Furthermore, for $A\subseteq Paths(M)$, define $Po^M(A)=\vee\{Po^M(\pi)|\pi\in A\}$, then we have a well-defined function $Po^M: 2^{Paths(M)}\longrightarrow[0,1].$ We call $Po^M$ the possibility measure over $\Omega=2^{Paths(M)}$ as it satisfies the conditions (i)-(iii) in Definition \ref{sec1:algebra} as shown in Theorem \ref{sec1:theo3} below. If $M$ is clear form the context, then $M$ is omitted and we simply write $Po$ for $Po^M$.}
\end{Definition}	

      First, let us see how to calculate the possibility measure $Po$ over the cylinder sets.

     \begin{proposition}\label{sec1:pro possmeas}
     Let $M$ be a possibilistic Kripke structure. Then the possibility measure of the cylinder sets are given by
\begin{equation*}
Po(Cyl(s_0\cdots s_n))=I(s_0)\wedge\bigwedge_{i=0}^{n-1}P(s_i,s_{i+1}),
\end{equation*}
where $P(s_0)=1$, specially, $Po(Cyl(s_0))=I(s_0)$.
\end{proposition}

     \begin{proof}
    As $Cyl(s_0\cdots s_n)=\cup\{\pi\in S^M|s_0\cdots s_n\in Pref(\pi)\}$, then
\begin{eqnarray*}
&&Po(Cyl(s_0\cdots s_n))\\
&=&\vee\{Po(\pi)|s_0\cdots s_n\in Pref(\pi)\}\\
&=&\vee\{I(s_0)\wedge\bigwedge_{i=0}^{\infty} P(s_i,s_{i+1}) | s_0\cdots s_n \in Pref(\pi), s_{n+1},  \cdots\in S\}\\
&=&\{I(s_0)\wedge\bigwedge_{i=0}^{n-1} P(s_i, s_{i+1})\}\wedge\vee\{\bigwedge_{i=n}^\infty P(s_i,s_{i+1})| s_i\in S,  i> n\}.
  \end{eqnarray*}

          Since $P$ is a possibilistic distribution, $i.e.$, for all states $s$, $\vee_{s'\in S} P(s,s')=1$  holds. Then, for any
  $\varepsilon>0,s\in S$, there exists $t\in S$, such that $P(s,t)\geq1-\varepsilon$. It follows that for all non-negative integer $i$, if $i\geq n$, then there exists $s_i\in S$ such that $P(s_i,s_{i+1})\geq1-\varepsilon$. This shows \begin{equation*}\vee\{\bigwedge_{i=n}^\infty P(s_i,s_{i+1})|s_i\in S\}=1.
  \end{equation*} So
  \begin{equation*}
  Po(Cyl(s_0\cdots s_n))=I(s_0)\wedge\bigwedge_{i=0}^{n-1}P(s_i,s_{i+1}).
  \end{equation*}
\end{proof}

     In the following, we also use $Po(s_0\cdots s_n)$ to represent $\wedge_{i=0}^{n-1}P(s_i,s_{i+1})$.

      \begin{theorem}\label{sec1:theo3}
     $Po$ is a possibility measure on $\Omega=2^{Paths(M)}$.

	In general, the following condition does not hold:

	If $A_1\supseteq A_2\supseteq\cdots$is a decreasing chain in $\Omega$ , then $Po(\cap_{i\in I}A_i)=\wedge_{i\in I}Po(A_i)$.

\end{theorem}

   \begin{proof} Let us show that $Po$ satisfies conditions (i)-(iii) in Definition \ref{sec1:algebra}.

   (i) (1) As $Po(\emptyset)=\vee\{Po(\pi)|\pi\in\emptyset\}=0$, then $Po(\emptyset)=0$.

(2)As $Paths(M)=\cup_{s\in S}Paths(s)=\cup_{s\in S}Cyl(s)$, then $Po(Paths(M))=\vee_{s\in S}Po(Cyl(s))$, according to Proposition \ref{sec1:pro possmeas}, $Po(Paths(M))=\vee_{s\in S}I(s)$. Since $I$ is a possibilistic distribution, $i.e$.,  $\vee_{s\in S}I(s)=1$. Hence, $Po(Paths(M))=1$.

(ii) Since $Po(\cup_{i\in I}A_i)=\vee\{Po(\pi)|\pi\in\cup_{i\in I}A_i\}=\vee_{i\in I}\{Po(\pi)|\pi\in A_i\}=\vee_{i\in I} Po(A_i)$. The condition (ii) holds.

(iii)  holds trivially by the definition of $Po$.

 In Fig.2, let $\pi_i=s_i^is_2^w$, $A_i=\{\pi_i,\pi_{i+1},\cdots\}$, then $\cap_{i\in I}A_i=\emptyset$, and thus $Po(\cap_{i\in I}A_i)=Po(\emptyset)=0$. On the other hand, $Po(\pi_i)=Po(s_i^is_2^w)=1/2$, then we get  $Po(A_i)=\vee\{Po(\pi_i),Po(\pi_{i+1}),\cdots\}=1/2$, it follows that $\wedge_{i\in I}Po(A_i)=1/2$. Hence, $Po(\cap_{i\in I}A_i)\neq\wedge_{i\in I}Po(A_i)$
\end{proof}

\begin{figure}[ht]
\begin{center}
\includegraphics[scale=0.5]{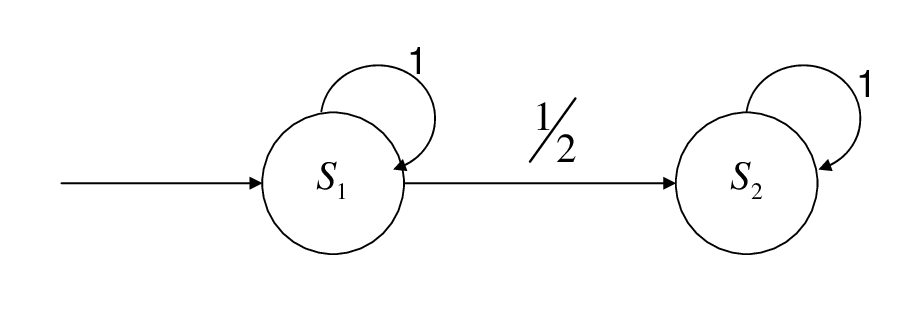}
\center{Fig.2.}A possibilistic Kripke structure $M$ with two states
\vspace{-0.3cm}
\end{center}
\end{figure}

       \begin{Remark}
     {\rm  (1) For path starting in a certain (possibly noninitial) state $s$, the same construction is applied to the possibilistic Kripke structure
 $M_s$ that resulting from $M$ by letting $s$ as the unique initial state. Formally, for $M=(S,P,I,AP,L)$ and state $s$,  $M_s$ is defined by $M_s=(S,P,s,AP,L)$ , where $s$ denotes an initial distribution with only one initial state $s$.

     (2) The possibility measure $Po$ is defined on the algebra $\Omega=2^{Paths(M)}$. Of course it can also be restricted to a
 $\sigma-$algebra $\Omega_\sigma$. The latter case corresponds with the probabilistic model checking (\cite{CJP08}). Clearly in possibilistic Kripke structure, the possibility measure is defined on $\Omega=2^{Paths(M)}$, thus for any set $A\in 2^{Paths(M)}$ , $A$ can be measured in the sense of $Po$. And then we can make the discussion more widely compared with probability measure.}
\end{Remark}

     \begin{example}
   {\rm  Consider the possibilistic Kripke structure $M$ in Figure 1, let us give several calculations of possibility measure.

$Po(Cyl(s_0s_1s_2))=I(s_0)\wedge P(s_0,s_1)\wedge P(s_1,s_2)=1$.

$Po(Cyl(s_0s_1s_2)^c)=Po(Cyl(s_0s_2)\cup Cyl(s_0s_1s_3))=Po(Cyl(s_0s_2))\vee Po(Cyl(s_0s_1s_3))$,
and since $Po(Cyl(s_0s_2))=I(s_0)\wedge P(s_0,s_2)=0.2$, $Po(Cyl(s_0s_1s_3))=I(s_0)\wedge P(s_0,s_1)\wedge P(s_1,s_3)=0.9$,
then $Po(Cyl(s_0s_1s_2)^c)=0.2\vee0.9=0.9$. It follows that $Po(Cyl(s_0s_1s_2)^c)+Po(Cyl(s_0s_1s_2))\neq1$.}
\end{example}

      It follows that possibility measures are not additive, and hence are different from probability measures. Model-checking based on a possibility
measure has different behavior compared with that based on a probability measure. We refer to \cite{Dra95} for further comparison between probability measures and possibility measures.

\section{Possibility  measures of reachability and repeated reachability linear-time properties}\label{sec4}

      The quantitative model-checking problem that we are confronted with is: given a possibilistic Kripke structure $M$ and an $LT$ property
 $P$, compute the possibility measure for the set of paths in $M$ for which $P$ holds. We consider some special cases: properties of reachability, constraint reachability and repeated reachability, in this section. For this purpose, let us first present the notion of linear-time properties in a possibilistic Kripke structure.

 \subsection{Linear-time properties}\label{sec1:Linear}

      Some of the relevant definition of $LTL$ (short for {\sl linear-temporal logic}) are presented as follows:

   \begin{Definition}
    (\cite{CJP08})(Syntax of $LTL$) {\rm $LTL$ formulae over the set $AP$ of atomic propositions are formed according to the following grammar,
\begin{equation}
\varphi ::=true | a|\varphi_1\wedge\varphi_2| \neg\varphi | \bigcirc\varphi|\varphi_1\cup\varphi_2
\nonumber
\end{equation}
where $a\in AP$.}
\end{Definition}

    For example, $\varphi=\bigcirc(\neg b\cup c)\wedge a$ is an $LTL$ formula, but $\varphi=\forall\bigcirc(b\vee c)$ is not, where $a,b,c\in AP$.

      \begin{Definition}(\cite{CJP08})
   (Semantics of $LTL$) {\rm Assume $\pi=s_0s_1s_2\cdots$ is a path starting $s_0$ in a possibilistic Kripke structure $M$, $\pi[i]=s_is_{i+1}s_{i+2}\cdots$, $a\in AP$
and set satisfaction relation $(\models)$ as follows:
\begin{eqnarray*}
\pi\models true;& \\
\pi\models a  & {\rm iff} \ a\in L(s_0);\\
\pi\models\varphi_1\wedge\varphi_2 & {\rm iff}\ \pi\models\varphi_1 \ {\rm and} \ \pi\models\varphi_2;\nonumber\\
\pi\models\neg\varphi  & {\rm iff} \ \pi\not\models\varphi;\\
\pi\models\bigcirc\varphi  & {\rm iff} \ \pi[1]\models\varphi;\\
\pi\models\varphi_1\cup\varphi_2  & {\rm iff} \  \exists k\geq0,\pi[k]\models\varphi_2
\ {\rm and} \\
&\ \ \pi[1]\models\varphi_1 {\rm \ for\ all} \ 1\leq i\leq k-1.\\
\end{eqnarray*}}
\end{Definition}

       The until operator allows to derive the temporal modalities $\lozenge$ (``eventually'', sometimes in the future) and  $\square$
 (``always'', from now on forever) as follows:
\begin{eqnarray*}
\lozenge\varphi=true\cup\varphi,\square\varphi=\neg\lozenge\neg\varphi.
\end{eqnarray*}

	As a result, the following intuitive meaning of $\lozenge$ and $\square$  is obtained. $\lozenge\varphi$ ensures that $\varphi$ will be true eventually in the future. $\square\varphi$ is satisfied if and only if it is not the case that eventually $\neg\varphi$ holds. This is equivalent to the fact that $\varphi$ holds from now on forever.

    \begin{Definition}
    {\rm Let $M=(S,P,I,AP,L)$ be a possibilistic Kripke structure without terminal states, $i.e$., for any state $s$, there exists a state $t$
such that $P(s,t)>0$. The trace of the infinite path fragment $\pi=s_0s_1\cdots$ is  defined as $trace(\pi)=L(s_0)L(s_1)\cdots$. The trace of the finite path fragment $\hat{\pi}=s_0s_1\cdots s_n$ is defined as  $trace(\hat{\pi})=L(s_0)L(s_1)\cdots L(s_n)$.}
\end{Definition}

 The set of traces of a set $\Pi$  of paths is defined in the usual way, $trace(\Pi)=\{trace(\pi)|\pi\in\Pi\}$. Let $Traces(s)$ denote
the set of traces initialed at $s$, and $Traces(M)$ the set of traces of the possibilistic Kripke structure $M$, $i.e$., $Traces(s)=trace(Paths(s))$  and $Traces(M)=\cup_{s\in S}Traces(s)$.

 $LTL$ formulae stand for properties of paths (or in fact their traces). This means that a path can either fulfill an $LTL$ formula or not.
To precisely formulate when a path satisfies an $LTL$ formula, we proceed as follows. First, the semantics of an $LTL$ formula $\varphi$ is defined as a language  $Words(\varphi)$ that contains all infinite words over the alphabet $2^{AP}$ which are traces of paths that satisfy $\varphi$. That is, $Words(\varphi)=\{trace(\pi)\in(2^{AP})^\omega|\pi\models\varphi\}$.

  \begin{Definition} {\rm A linear-time property ($LT$ property) over the set of atomic propositions $AP$ is a subset of $(2^{AP})^\omega$.}
\end{Definition}

        Note that it suffices to consider infinite words only (and not finite words), as possibilistic Kripke structure without terminal
states are considered.

    \begin{Definition}
  {\rm  Let $P$ be an $LT$ property over $AP$ and $M=(S,P,I,AP,L)$ be a possibilistic Kripke structure without terminal states. Then,
$M=(S,P,I,AP,L)$  satisfies $P$, denoted  $M\models P$, iff $Traces(M)\subseteq P$. State $s\in S$ satisfies $P$, notation $s\models P$, whenever $Traces(s)\subseteq P$.}
\end{Definition}

\subsection{Possibility measure of LT property}

        For a countable set $\Sigma$, any subset of $\Sigma^\omega$ is called a language of infinite words, sometimes also called an $\omega-$language. Languages will
 be denoted by the symbol $L$.

\begin{Definition}
     {\rm Let $M$ be a possibilistic Kripke structure and $P$ an $LT$ property (both over $AP$). The possibility for $M$ to exhibit a trace
in $P$, denoted $Po^M(P)$,  is defined by
\begin{eqnarray*}
Po^M(P)=Po^M(\{\pi\in Paths(M)|trace(\pi)\in P\}).
\end{eqnarray*}}
\end{Definition}

      Similarly, for the $LTL$ formula $\varphi$ we write $Po(\varphi)$ for $Po(Words(\varphi))$,
i.e., $Po(\varphi)=Po(Words(\varphi))=Po(\{\pi\in Paths(M)|\pi\models\varphi\})$. For state
$s$ of $M$, we write $Po(s\models\varphi)$ for $Po^{M_s}(\{\pi\in Paths(s)|\pi\models\varphi\})$, $i.e.$, $Po(s\models\varphi)=Po^{M_s}(\{\pi\in Paths(s)|\pi\models\varphi\})$.

\subsection{Reachability possibility}

     One of the elementary questions for the quantitative analysis of systems modeled by possibilistic Kripke structures is to compute
the possibility of reaching a set $B$ of states, where $B$ may represent a set of certain bad states which should be visited only with some small possibility, or dually, a set of good states which should rather be visited frequently.

    This subsection focuses on computing $Po(\lozenge B)$.

    The possibility measure of eventually reaching $B$ is given by:
\begin{eqnarray*}
Po(\lozenge B)&=&\vee_{s_0\cdots s_n\in Paths_{fin}(M)\cap (S\backslash B)^*B}Po(Cyl(s_0\cdots s_n))\\
&=&\vee_{s_0\cdots s_n\in Paths_{fin}(M)\cap (S\backslash B)^*B}\{I(s_0)\wedge\bigwedge_{i=1}^{n-1}P(s_i,s_{i+1})\}.
\end{eqnarray*}

Using the transitive closure $P^+$, $Po(\lozenge B)$ has a very simple form, $i.e$.,

$Po(\lozenge B)=\vee_{s\in S}\vee_{t\in B}I(s)\wedge P^+(s,t)$.\hfill (2)

In the following, we give another approach to calculate $P(\lozenge B)$, which is adopted in the probabilistic model-checking of reachability (\cite{CJP08}).

       Let variable $x_s$ denote the possibility measure of reaching $B$ from $s$, $i.e$., $x_s=Po(s\models\lozenge B)$, for arbitrary $s\in S$.
 The goal is to compute $x_s$ for all state $s$. There are three cases to consider.

(1) $B$ is not reachable from $s$ in the underlying directed graph of $M$, then $x_s=0$.

(2) $B$ is reachable from $s$, $i.e$., $x_s>0$. Moreover,  if $s\in B$, then $x_s=1$ . For the state $s$ is not in $B$ , it holds that
\begin{eqnarray*}
x_s=\vee_{t\in S\backslash B}\{P(s,t)\wedge x_t\} \vee \{\vee_{u\in B} P(s,u)\}.
\end{eqnarray*}

      This equation states that either $B$ is reached within one step, $i.e$., by a finite path fragment $su$ with $u\in B$ (second summand, $\vee_{u\in B} P(s,u)$),
or first a state $t\in S\backslash B$ is reached from which $B$ is reached - this corresponds to path fragments $st\cdots u$ of length $\geq2$ where all states (except the last one) do not belong to $B$ (first summand, $\vee_{t\in S\backslash B}\{P(s,t)\wedge x_t\}$). Let $\hat{S}=Pre^*(B)\backslash B$ denote the set of states $s\in S\backslash B$  such that there is a path fragment $s_0\cdots s_n(n>0)$ with $s_0=s$.   Then for the vector $X=(x_s)_{s\in \hat{S}}$, we have:

$X=A\circ X\vee b$, \hfill (3)

\noindent where $A=(P(s,t))_{s,t\in \hat{S}}$ and the vector $b=(b_s)_{s\in \hat{S}}$ contains the possibilities of reaching $B$ from $\hat{S}$ within one step, $i.e$., $b_s=Po(s,B)$.

       The above technique yields the following two phase algorithm to compute reachability possibility in finite Markov chains: first,
perform a graph analysis to compute the set $\hat{S}$ of all states that can reach $B$ ($e.g$., by backward $DFS$ (depth-first search) - or $BFS$ (width-first search) - based search from $B$), then generate the matrix $A$ and the vector $b$. Second, solve the fuzzy relation equation $X=A\circ X\vee b$. This problem is addressed below by characterizing the desired possibility vector as the least solution in $[0,1]^{\hat{S}}$. This characterization enables us to compute the possibility measure by a finite iteration method. In fact, we present a characterization for a slightly more general problem, viz. constrained reachability ($i.e$., until properties).

\subsection{Constrained reachability possibility}

	Let $M=(S,P,I,AP,L)$ be a possibilistic Kripke structure and $B,C\subseteq S$. Consider the event of reaching $B$ via a finite path fragment which ends in a state $s\in B$, and visits only states in $C$ prior to reaching $s$. Using $LTL-$like notations, this event is denoted by $C\cup B$. The event $\lozenge B$ considered above agrees with $S\cup B$. For $n\geq0$, the event
$C\cup^{\leq n} B$ has the same meaning as $C\cup B$, except that it is required to reach $B$ (via states in $C$) within $n$ steps. Formally, $C\cup^{\leq n} B$ is the union of the basic cylinders spanned by path fragments $s_0\cdots s_k$ such that $k\leq n$ and  $s_i\in C$ for all $0\leq i< k$ and $s_k\in B$.

      Let $S_{=0}, S_{=1}, S_{?}$ be a partition of $S$ such that,

 (1) $B\subseteq S_{=1} \subseteq\{s\in S| Po(s\models C\cup B)=1\}$;

 (2) $S\backslash C\cup B\subseteq S_{=0} \subseteq\{s\in S| Po(s\models C\cup B)=0\}$;

(3) $S_{?}=S\backslash(S_{=1}\cup S_{=0})$.

      For all state $s$, if $s\in S_{=1}$, we have $Po(s\models C\cup B)=1$; if $s\in S_{=0}$, $Po(s\models C\cup B)=0$; if $s\in S_{=?}$, we can get a
fuzzy matrix $A=(P(s,t))_{s,t\in S_{?}}$ by omitting the rows and columns for the states $s\in S_{=1}\cup S_{=0}$ from $P$ and $b=(b_s)_{s\in S_{?}}$.

     \begin{theorem}\label{sec3:Fixed}(Least Fixed Point Characterization)

  (i) The vector $X=(Po(s\models C\cup B))_{s\in S_?}$ is the least fixed point of the operator $\Phi:[0,1]^{S_?}\longrightarrow [0,1]^{S_?}$, which is given by $\Phi(Y)=A\circ Y\vee b$.

  (ii) Furthermore, if $X^{(0)}=0$ is the vector consisting of zeros only, and for $n\geq0$, $X^{(n+1)}=\Phi(X^n)$ , then

 -(ii-1) $X^{(n)}=(x_s^{(n)})_{s\in S_?}$, where $x^{(n)}_s=Po(s\models C\cup^{\leq n}S_{=1})$ for each state $s\in S_?$,

 -(ii-2) $X^{(0)}\leq X^{(1)} \leq X^{(2)}\cdots \leq X$,

 -(ii-3) $X=lim_{n\longrightarrow\infty}X^{(n)}$, where $X=lim_{n\longrightarrow\infty}X^{(n)}$ means that $x_s=lim_{n\longrightarrow\infty}x_s^{(n)}$ in the usual sense for any $s\in {S_?}$.
 \end{theorem}

       \begin{proof}
   (i) We prove the well-definiteness of $\Phi$ as a function from $[0,1]^{S_?}$ to $[0,1]^{S_?}$. For $Y=(y_s)_{s\in S_?}$,
the vector $\Phi(Y)=(y'_s)_{s\in S_?}$ and $y'_s=\vee_{t\in S_?}\{P(s,t)\wedge y_t\}\vee P(s,S_{=1})$. Clearly, $y'_s\in[0,1]^{S_?}$ , therefore $\Phi(Y)\in[0,1]^{S_?}$.

      Next we prove the fixed point property,  $i.e$., $X=\Phi(X)$ .

 Since if $s \in S_{=1}$, then $x_s=Po(s\models C\cup B)=1$; if $s \in S_{=0}$, then $x_s=Po(s\models C\cup B)=0$. For all
  $s \in S_{?}$, we derive that,
 $x_s=\vee_{t\in S}(P(s,t)\wedge x_t)
 =\{\vee_{t\in S_{=0}}(P(s,t)\wedge x_t)\}\vee\{\vee_{t\in S_?}(P(s,t)\wedge x_t)\}
 \vee\{\vee_{t\in S_{=1}}(P(s,t)\wedge x_t)\}
 =0\vee\{\vee_{t\in S_?}(P(s,t)\wedge x_t)\}\vee\{\vee_{t\in S_{=1}}(P(s,t)\wedge x_t)\}
 =\{\vee_{t\in S_?}(P(s,t)\wedge x_t)\}\vee P(s,S_{=1})$,
\noindent which is the component for state $s$ in the vector $\Phi(X)$. Hence $X=\Phi(X)$.

    (ii) Let us first show that $x_s^{(n)}=Po(s\models C\cup^{\leq n} S_{=1})$ for each state $s\in S_?$ by induction on $n$.

If $n=0, x_s^{(0)}=0$, and $Po(s\models C\cup^{\leq n} S_{=1})=Po(\varnothing)$, so $x_s^{(0)}=Po(s\models C\cup^{\leq n} S_{=1})$.

If $n=1$, \begin{eqnarray*}
x^{(1)}_s=\{\vee_{t\in S_?}(P(s,t)\wedge x^{(0)}_t\}\vee P(s,S_{=1})
=\vee_{u\in  S_{=1}}P(s,u)
\end{eqnarray*} and
\begin{eqnarray*}
Po(s\models C\cup^{\leq 1}s_{=1})&=&\vee_{s_1\in S_{=1}}Po(Cyl(ss_1))
=\vee_{s_1\in S_{=1}}P(ss_1)\\
&=&\vee_{s_1\in S_{=1}}P(s,s_1)
=\vee_{u\in S_{=1}}P(s,u).\end{eqnarray*}

 Hence, $x_s^{(1)}=Po(s\models C\cup^{\leq 1}S_{=1})$.

 Assume that for any $n\geq2, x^{(n)}_s=Po(s\models C\cup^{\leq n}S_{=1})$, let us calculate $x^{(n+1)}_s$ as follows:
\begin{eqnarray*}
x^{(n+1)}_s &=& \Phi(x^{(n)}_s)\\
 &=&\vee_{t_0\in S_?}(P(s,t_0)\wedge x^{(n)}_s)\vee P(s,S_{=1})\\
 &=&\{\vee_{t_0\in S_?}P(s,t_0)\wedge Po(t_0\models C \cup^{\leq n} S_{=1})\} \\ & &\vee P(s,S_{=1})\\
 &=&\vee_{t_0\in S_?}P(s,t_0)\wedge \vee\{I(t_0)\wedge P(t_0\cdots t_k)|\ k\leq n, \\
 & &  t_k\in S_{=1} \ {\rm and \  for  \  all}\ 0\leq i \leq k, \ t_i\in C\}\\ & & \vee P(s,S_{=1})\\
 &=&\vee_{t_0\in S_?}P(s,t_0)\wedge(\vee_{t_0\in C,t_1\in S_{=1}}(I(t_0)\wedge P(t_0t_1))\\
 &=&\vee_{t_0,t_1\in C,t_2\in S_{=1}}(I(t_0)\wedge P(t_0t_1t_2))\vee\cdots\vee\\ &&\ \ \bigvee_{t_0,\cdots, t_{n-1}\in C,t_n\in S_{=1}}(I(t_0)\wedge P(t_0t_1t_2\cdots t_n))\\ & & \vee P(s,S_{=1})\\
 &=&\vee\{Po(Cyl(st_0t_1t_2\cdots t_k))| \ k\leq n, t_k\in S_{=1} \\ & & {\rm and\   for  \ all  \ } 0\leq i \leq k, \ t_i\in C\}\\
 &=& Po(s \models C\cup^{\leq n+1} S_{=1}).
\end{eqnarray*}
Hence,\begin{eqnarray*}
x_s^{(n+1)}=Po(s\models C\cup^{\leq n+1}S_{=1}).
\end{eqnarray*}
This shows that $x_s^{(n)}=Po(s\models C\cup^{\leq n}S_{=1})$ for each state $s\in S_?$, and the condition (ii-1) holds.

Since $C\cup S_{=1}$ is the countable union of the events $C\cup^{\leq n} S_{=1}$,  for $s\in S_?$, let $x'_s=Po(s\models C\cup S_{=1})$, then we have,
\begin{eqnarray*}
x'_s&=&Po(\cup^\infty_{n=1}(s\models C\cup^{\leq n}S_{=1}))\\
&=&lim_{n\longrightarrow \infty} Po(s\models C\cup^{\leq n}S_{=1})\\
&=&lim_{n\longrightarrow \infty}x_s^{(n)}.
\end{eqnarray*}
then $X'=lim_{n\longrightarrow\infty}X^{(n)}$, where $X'=(x'_s)_{s\in S_?}$. Since $B\subseteq S_{=1}$, it follows that
$x'_s=Po(s\models C\cup S_{=1})\geq Po(s\models C\cup S)=x_s$, and thus $X'\geq X$.

 Now we prove $X^{(n)}\leq X^{(n+1)}\leq X$ by induction on non-negative integer number $n$. Since $X^{(0)}=0$,
 $X^{(0)}=0\leq b=A\circ 0\vee b=X^{(1)}=\Phi(0)\leq \Phi(X)=X$. Assume that $X^{(n-1)}\leq X^{(n)}\leq X$ for $n\geq1$, then
 $X^{(n)}=A\circ X^{(n-1)}\vee b \leq A\circ X^{(n)}\vee b=X^{(n+1)}= \Phi(X^{(n)})\leq \Phi(X)=X$. It follows that $X^{(0)}\leq X^{(1)}\leq X^{(2)}\leq \cdots \leq lim_{n\longrightarrow\infty}X^{(n)}=X^{\prime}\leq X$.
Hence $lim_{n\longrightarrow\infty}X^{(n)}=X^{\prime}=X$. The conditions (ii-2) and (ii-3) hold.

 Finally, we prove that $X=(Po(s\models C\cup B))_{s\in S_?}$ is the least fixed point. If $Y$ is an arbitrary fixed point of $\Phi(X)=A\circ X\vee b$, since $0\leq Y$, and $X^{(1)}=\Phi(X^{(0)})\leq \Phi(Y)=Y$, $X^{(2)}=\Phi(X^{(1)})\leq \Phi(Y)=Y$, $X^{(n)}=\Phi(X^{(n-1)})\leq \Phi(Y)=Y$, $\cdots$, it follows that $X=lim_{n\longrightarrow\infty}X^{(n)}\leq Y$. Therefore, $X=(Po(s\models C\cup B))_{s\in S_?}$ is the least fixed point of the operator $\Phi$.
\end{proof}

    \begin{Remark} {\rm For a finite Kripke structure $M$, the fuzzy matrix $A$ and $b$ are finite. Since the operations involved in the iteration
of the least fixed point characterization in Theorem \ref{sec3:Fixed} are maximum and minimum operations over the unit interval [0,1], it follows that the iteration will end after a finite number of steps. In fact, in $|S_?|$ steps, this iteration will end, where $|S_?|$ denotes the number of states in $S_?$. Therefore, we can effectively compute the exact solution in Theorem \ref{sec3:Fixed}.}
\end{Remark}

       \begin{example}{\rm	Consider the possibilistic Kripke structure $M$ in Example \ref{sec1:ep1}, the event of interest is
$C\cup^{\leq n} B$ where $B=\{s_3\}$, $C=\{s_0,s_1,s_2\}$. We shall compute the bounded constrained reachability possibility $x_s=Po(s\models C\cup^{\leq n} B)_{s\in S_?}$for all states $s\in S_?$, where we take $S_0=\emptyset$, $S_{=1}=\{s_3\}$, $S_{?}=\{s_0,s_1,s_2\}$.

     Using the state order $s_0< s_1<s_2$, the possibility matrix $A$ and the vector $b$ are given by,

$A=\left(\begin{array}{cccc}
0&1&0.2\\
0&0&1\\
0&0.7&0
\end{array}
\right)$,$b=\left(\begin{array}{cccc}0\\
0.9\\
1
\end{array}
\right)$.

      The least fixed point characterization suggests the following iterative scheme,

       $X^{(0)}=0$ and $X^{(n+1)}=A\circ X^{(n)}\vee b$,

\noindent where $X^{(n)}=(Po(s\models C\cup^{\leq n}S_{=1}))_{s\in S_?}$. Then we can obtain that
\begin{eqnarray*}
X^{(1)}&=&A\circ X^{(0)} \vee b=b,\\
X^{(2)}&=&A\circ X^{(1)} \vee b\\
&=&\left(\begin{array}{cccc}
0&1&0.2\\
0&0&1\\
0&0.7&0
\end{array}
\right)\circ\left(\begin{array}{cccc}0\\
0.9\\
1
\end{array}
\right)\vee\left(\begin{array}{cccc}0\\
0.9\\
1
\end{array}
\right)\\
&=&\left(\begin{array}{cccc}0.9\\
1\\
1
\end{array}
\right),\\
X^{(3)}&=&A\circ X^{(2)} \vee b\\
&=&\left(\begin{array}{cccc}
0&1&0.2\\
0&0&1\\
0&0.7&0
\end{array}
\right)\circ\left(\begin{array}{cccc}0.9\\
1\\
1
\end{array}
\right)\vee\left(\begin{array}{cccc}0\\
0.9\\
1
\end{array}
\right)\\
&=&\left(\begin{array}{cccc}1\\
1\\
1
\end{array}
\right).
\end{eqnarray*}

It follows that $X^{(n)}=X^{(3)}$ for any $n\geq3$. That is,
\begin{eqnarray*}
X&=&(Po(s\models C\cup S_{=1}))_{s\in S_?}\\
&=&lim_{n\longrightarrow\infty}X^{(n)}\\
&=&\left(\begin{array}{cccc}1\\
1\\
1
\end{array}
\right).
\end{eqnarray*}
	
      On the other hand, clearly, $X=(Po(s\models C\cup S_{=1}))_{s\in S_?}=(Po(s\models \lozenge B))_{s\in \hat{S}}$. Using Theorem \ref{sec3:Fixed} to
the fuzzy relation equation
	\begin{eqnarray*}
X=A\circ X\vee b
\end{eqnarray*} where $X=\left(\begin{array}{cccc}x_{s_0}\\
x_{s_1}\\
x_{s_2}
\end{array}
\right)$, this equation can be rewritten as
\begin{eqnarray*}
\left(\begin{array}{cccc}x_{s_0}\\
x_{s_1}\\
x_{s_2}
\end{array}
\right)=\left(\begin{array}{cccc}0&1&0.2\\
0&0&1\\
0&0.7&0
\end{array}
\right)\circ \left(\begin{array}{cccc}x_{s_0}\\
x_{s_1}\\
x_{s_2}
\end{array}
\right)\vee\left(\begin{array}{cccc}0\\
0.9\\
1
\end{array}
\right),
\end{eqnarray*}
solving this fuzzy relation equation yields the solution
$X=\left(\begin{array}{cccc}1\\
1\\
1
\end{array}
\right)$.}
\end{example}

\subsection{Repeated reachability possibility}

      This section focuses on quantitative properties of repeated reachability of finite possibilistic Kripke structure which can
be verified using graph analysis, $i.e$, by just considering the underlying digraph of the finite possibilistic Kripke structure, combining the transition possibility distribution.
	
For a finite possibilistic Kripke structure $M$, let $B\subseteq S$ be a set of states in $M$, and $s$ a state in $M$.  For the event $\square\lozenge B$, $i.e$., the set of all paths that visit $B$ infinitely, let us calculate $Po(s\models\square\lozenge B)$. Since $Po(s\models\square\lozenge B)=\vee_{a\in B}Po(s\models\square\lozenge a)$, it suffices to calculate the possibility $Po(s\models\square\lozenge a)$ for any $a\in B$. Let us give some analysis on how to calculate $Po(s\models\square\lozenge a)$ in the following.

\begin{theorem}
      For a finite possibilistic Kripke structure $M$, $a\in S$, we have

$Po(s\models\square\lozenge a)=\vee_TD_T$, \hfill (4)

\noindent where $T$ ranges over all strongly connected subsets of the diagraph of $M$ such that $a\in T$,  and for $T=\{t_1,\cdots,t_k\}$,

$D_T=\vee_{\varphi\in S_k}\wedge_{i=0}^kP^+(t_{\varphi(i)},t_{\varphi(i+1)})$, \hfill (5)

\noindent where $S_k$ denotes the set of all permutations on the set $\{1,\cdots,k\}$ and $t_{\varphi(0)}=s,t_{\varphi(k+1)}=t_{\varphi(1)}$.
\end{theorem}

    Recall a strongly connected subset of the diagraph $M$ is a subset $T$ of $S$ such that, for any $s,t\in T$ ,
there are state sequence $t_1\cdots t_k$ in $T$ such that $t_1=s,t_k=t$ and $P(t_i,t_{i+1})>0$ for any $1\leq i\leq k-1$.

  \begin{proof}
    Let $T=\{t_1,\cdots,t_k\} $ be a strongly connected subset of $M$ containing $a$. For $\varphi\in S_k$,
if $\wedge_{i=0}^kP^+(t_{\varphi(i)},t_{\varphi(i+1)})\neq0$, by the observation in Remark \ref{sec1:remark2}, for any $i$, there exists a state sequence $m(t_{\varphi(i)},t_{\varphi(i+1)})$ that leads from state $t_{\varphi(i)}$ to  state $t_{\varphi(i+1)}$ such that $P^+(t_{\varphi(i)},t_{\varphi(i+1)})=P(m(t_{\varphi(i)},t_{\varphi(i+1)}))$. Let $\pi=m(s,t_{\varphi(1)})(m(t_{\varphi(1)},t_{\varphi(2)})\cdots $ $m(t_{\varphi(k-1)},t_{\varphi(k)}))^\omega$. 	 Since $a\in T$, it follows that $\pi\models\square\lozenge a$, and $Po^{M_s}(\pi)=\wedge_{i=0}^kP^+$ $(t_{\varphi(i)},t_{\varphi(i+1)})$. This shows that $D_T\leq Po(s\models\square\lozenge a)$ for any strongly connected subset $T$ of $M$ such that $a\in T$.

       Conversely,  if $\pi\models\square\lozenge a$ and the first state of $\pi$ is $\pi(1)=s$, let $T=inf(\pi)$, the set of states
 occurring infinitely often in $\pi$. Then  $T$ is a strongly connected subset of $M$ such that $a\in T$. Assume that  $T=\{t_1,\cdots,t_k\} $ , let $t_j=min\{n|\pi(n)=t_j\}$ for any $1\leq j\leq k$, where $\pi(n)$ denotes the $n$-th state of $\pi$. We may assume that $t_1< t_2 <\cdots t_k$. Let $\pi'=s(\pi(i_1)\cdots \pi(i_k))^\omega$. Since $\pi$  has the form $s\cdots t_1 \cdots t_2\cdots t_k\cdots t_1\cdots$, by the definition of $P^+$, it follows that $Po^{M_s}(\pi)\leq Po^{M_s^+}(\pi')$, where $Po^{M_s^+}(\pi')=\wedge^k_{j=0}P^+(t_{i_j},$ $t_{i_{j+1}})$. Hence, $Po^{M_s}(\pi)\leq D_T$. This shows that $Po(s\models\square\lozenge a)\leq\vee_T D_T$, where $T$ ranges over all strongly connected subset of the diagraph of $M$ such that $a\in T$.

Therefore, we have the required equality.
\end{proof}

     There may be many strongly connected subsets of $M$ containing $a$, they do not play the same role in the calculation
$Po(s\models\square\lozenge a)$ as shown by the following lemma.

\begin{lemma}
    For a finite possibilistic Kripke structure $M$, if  $T'$ and $T$ both are strongly connected subset of  $M$  containing $a$, and
$T'\subseteq T$, then we have $D_T\leq D_{T'}$.
\end{lemma}

     This is obvious by the definition of $D_T$ and $D_{T'}$.

      By this lemma, to calculate $Po(s\models\square\lozenge a)$, it is sufficient to calculate $D_T$ for those minimal strongly
connected subsets containing $a$ in $M$. Note that if $P^+(a,a)\neq0$, the minimal strongly connected subset containing $a$ in $M$ is unique, $i.e$., $T=\{a\}$. In this case,

$Po(s\models\square\lozenge a)=D_T=P^+(s,a)\wedge P^+(a,a)$. \hfill (6)

If $P^+(a,a)=0$, then there is no strongly connected subset containing $a$ in $M$. In this case, $Po(s\models\square\lozenge a)=0$, and the equality $Po(s\models\square\lozenge a)=P^+(s,a)\wedge P^+(a,a)$ also holds.

Therefore, in any case, $Po(s\models\square\lozenge a)=P^+(s,a)\wedge P^+(a,a)$. Then we have the following theorem.

\begin{theorem}\label{sec3:theor6}
       Let $M$ be a finite possibilistic Kripke structure and $B\subseteq S$. Then we have,

$Po(s\models\square\lozenge B)=\vee_{a\in B}P^+(s,a)\wedge P^+(a,a)$. \hfill (7)

\end{theorem}

    Since the calculation of $P^+$ can be done by some simple graph-search algorithm combining with the minimum and maximum operations
in the unit interval [0,1] or some simple fuzzy matrix algorithms, then $Po(s\models\square\lozenge B)$ can be effectively calculated.

     In the probabilistic model checking of repeated reachability linear-time properties (see Ref.\cite{CJP08}),
a different approach which is not appropriate to possibilistic model checking is adopted, which is more complex than our method for
the possibilistic model checking of repeated reachability linear-time properties.

\begin{example}
    {\rm   Consider the possibilistic Kripke structure $M$ in Example 1. By a simple calculation, the corresponding possibilistic Kripke
structure $M^+$  using the transitive closure $P^+$  as the transition possibility distribution is presented in Fig. 3. Then, by Theorem \ref{sec3:theor6}, we have $Po(s_0\models\square\lozenge s_1)=P^+(s_0,s_1)\wedge P^+(s_1,s_1)=0.7$, $Po(s_0\models\square\lozenge s_2)=P^+(s_0,s_2)\wedge P^+(s_2,s_2)=0.7$,  $Po(s_0\models\square\lozenge s_3)=P^+(s_0,s_3)\wedge P^+(s_3,s_3)=1$.
\begin{figure}[ht]
\begin{center}
\includegraphics[scale=0.5]{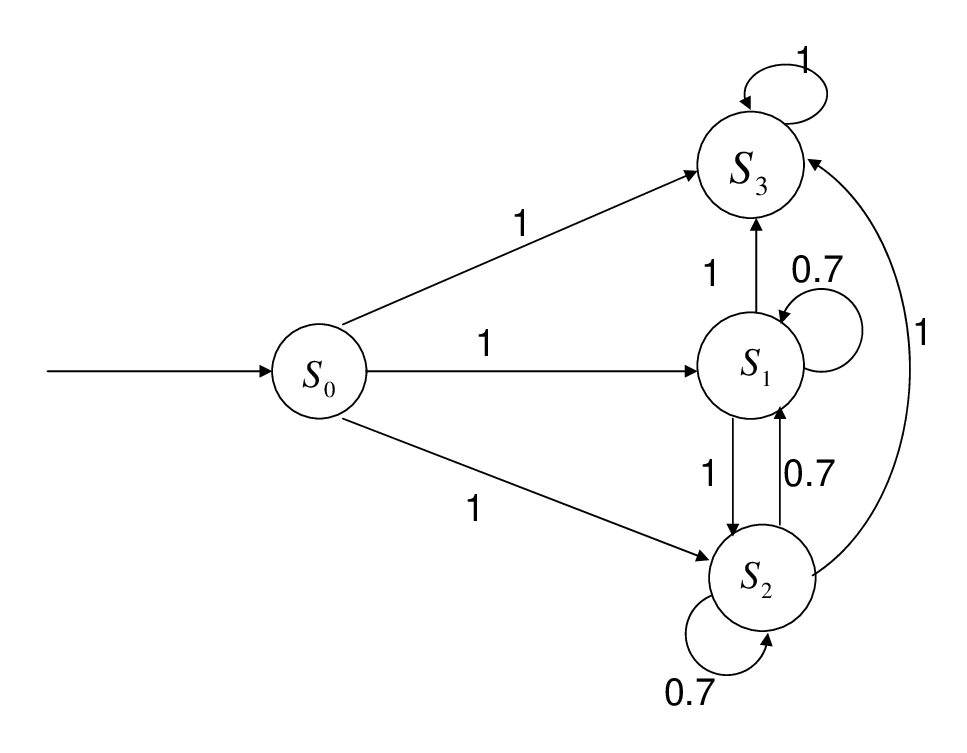}
\center{Fig.3.}	The corresponding $M^+$  of $M$ in Fig.1
\vspace{-0.3cm}
\end{center}
\end{figure}}
\end{example}

\section{Possibility measures of linear-time properties}\label{sec5}

       We continue to study more general linear-time properties in this section.

    Given an $\omega$-regular property $P$ (($c.f$. \cite{MOD59,MYP86,MYP94}) over $AP$ and a finite possibilistic Kripke structure $M=(S,P,I,AP,L)$, the goal is to compute $Po^M(P)$. The $LT$ property $P$ is represented by means of a finite automaton ${\cal A}$. The possibilistic Kripke structures in this section are assumed finite.

\subsection{Finite automata over finite words and infinite words}

     First, let us recall the notion of finite automata theory (\cite{CJP08,MOD59}).

       A (nondeterministic) finite automaton $(NFA)$  is a tuple ${\cal A}=(Q,\Sigma,\delta,I,F)$, where $Q$ denotes a finite set of states,
 $\Sigma$ is a finite input alphabet, $\delta:Q\times\Sigma\longrightarrow2^Q$ is a transition function, $I\subseteq Q$ is a set of initial states, and $F\subseteq Q$ is a set of accept (or final) states.

          The transition function $\delta$ can be identified with the relation $\longrightarrow\subseteq Q\times\Sigma\times Q$ given
  by $q\overset{u}{\longrightarrow} q'$ iff $q'\in\delta(q,u)$. Thus, often the notion of transition relation (rather than transition function) is used for $\delta$. Intuitively, $q\overset{u}{\longrightarrow} q'$ denotes that the automaton can move from state $q$ to state $q'$ when reading the input symbol  $u$.

      Next, we give the notion of a run for a finite automaton ${\cal A}$. Let $w=u_1\cdots u_n$ be a finite word. A run for $w$ in ${\cal A}$
is a finite sequence of states $q_0q_1\cdots q_n$ such that $q_i\overset{u_{i+1}}{\longrightarrow} q_{i+1}$ for all $0\leq i\leq n-1$.

      A run $q_0q_1\cdots q_n$ is called successful if $q_0\in I$ and $q_n\in F$. A finite word $w\in\Sigma^*$ is called accepted by ${\cal A}$
if there is a successful run for $w$. The accepted language of ${\cal A}$, denoted $L({\cal A})$ is the set of finite word in $\Sigma^*$ accepted by ${\cal A}$, $i.e$.
$L({\cal A})=\{w\in\Sigma^*|$ there  is a successful run
 in ${\cal A}$ for the word $w\}$. For a language $L\subseteq\Sigma^*$, if there is an NFA ${\cal A}$ such that $L=L({\cal A})$, then $L$ is called a regular language over $\Sigma$.

      When an $NFA$ is an acceptor of infinite word, then we have the notion of (nondeterministic) B\"{u}chi automaton $(NBA)$. An $NBA$
has the same structure as an $NFA$, ${\cal A}=(Q,\Sigma,\delta,I,F)$, say, the difference is the run and the accepting language of ${\cal A}$.

     Let $\sigma=u_1u_2\cdots\in\Sigma^\omega$ be an infinite word, a run for $\sigma$ in ${\cal A}$  is an infinite sequence of states $q_0q_1\cdots$
such that $q_i\overset{u_{i+1}}{\longrightarrow} q_{i+1}$ for all $i\geq0$. A run $q_0q_1\cdots$ is accepting if $q_0\in I$ and $q_i\in F$ for infinite many indices $i$. The accepting (infinite) language of ${\cal A}$ is

$L_\omega({\cal A})=\{\sigma\in\Sigma^\omega|$ there is a successful run  in ${\cal A}$  for  the  word  $\sigma\}.$

For a language $L\subseteq\Sigma^\omega$, if there is an $NBA$  ${\cal A}$ such that $L=L_\omega({\cal A})$, then $L$ is called an $\omega$-regular language over $\Sigma$.

\subsection{Possibility measure of regular safety property $P_{safe}$}

     Safety properties are often characterized as ``nothing bad should happen''. Formally, in classical case, safety property is defined as
an $LT$ property over $AP$ such that any infinite word $¦Ò$ where $P$ does not hold contains a bad prefix. For convenience, we use the dual notion of good prefixes to define safety property here. Of course, they are equivalent.

\begin{Definition}
    {\rm For a property $P$, the good prefixes of $P$, say $GPref(P)$ is defined by
\begin{eqnarray*}
GPref(P)=\{\hat{\sigma}\in(2^{AP})^*|\hat{\sigma}\in Pref(\sigma),\sigma\in P\},
\end{eqnarray*}}
\end{Definition}

      An $LT$ property $P_{safe}$ is called a safety property provided that, for a $\sigma$, if for all $\hat{\sigma}\in Pref(\sigma)$,
$\hat{\sigma}\in GPref(P_{safe})$, then $\sigma\in P_{safe}$, i.e.,

$\{\sigma\in(2^{AP})^\omega|\forall\hat{\sigma}\in Pref(\sigma),\hat{\sigma}\in GPref(P_{safe})\}=P_{safe}.$


\begin{Definition}
    {\rm  A safety property $P_{safe}$ over $AP$ is called regular if its set of good prefixes constitutes a regular language over $2^{AP}$.}
\end{Definition}

     For $regular$ safety property $P_{safe}$ there is an automaton accepting the good prefixes $GPref(P_{safe})$.

       Let ${\cal A}=(Q,2^{AP},\delta,I,F)$ be an $NFA$ for the good prefixes of a regular safety property $P_{safe}$. That is,
 $P_{safe}=\{A_0A_1\cdots (2^{AP})^\omega|$ for all $ n\geq 0, A_0A_1\cdots A_n\in L({\cal A})\}$ where $\delta(q,A)$ is defined for each $A\subseteq AP$ and each state $q\in Q$. Furthermore, let $M=(S,P,I,AP,L)$ be a finite possibilistic Kripke structure. Our interest is to compute the possibility $Po^M(P_{safe})=\vee_{s\in S}I(s)Po(s\models {\cal A})$ for $M$ to generate a trace in $P_{safe}$. The value $Po(s\models {\cal A})$ can be written as $Po(s\models {\cal A})=\vee_{\hat{\pi}}P(\hat{\pi})$ where $\hat{\pi}$ ranges over all finite path $s_0s_1\cdots s_n$ starting in $s_0=s$ such that $trace(s_0s_1\cdots s_n)=L(s_0)L(s_1)\cdots L(s_n)\in L({\cal A})$ and $Pref(trace(s_0s_1\cdots s_n))\subseteq GPref(P_{safe})$.

       Computing the values $Po(s\models {\cal A})$ by using these $\vee_{\hat{\pi}}P(\hat{\pi})$ may be difficult. Alternatively, we adapt the classical
 techniques for checking regular safety properties of transition systems to the possibilistic case. This involves the product of $M$ and ${\cal A}$ which is defined as follows.

\begin{Definition}
     {\rm    Let $M=(S,P,I,AP,L)$ be a possibilistic Kripke structure and ${\cal A}=(Q,2^{AP},\delta,I,F)$ be an $NFA$.
The product $M\otimes {\cal A}$ is a possibilistic Kripke structure, $M\otimes {\cal A}=(S\times Q,P',I',AP',L')$ , where

(1)
$AP'=S\times Q$, and $L'(\langle s,q\rangle)=\langle s,q\rangle$  for  any  $\langle s,q\rangle\in S\times Q$;

(2)
\begin{equation*}
I'(\langle s,q\rangle)=\left\{
\begin{array}{cc} I(s),& $if$\ q \in\delta (q_0,L(s))$ \ for some$\ q_0\in I;\\
0,& $otherwise$.
\end{array}
\right.
\end{equation*}

(3)
the transition possibility distribution of $M\otimes {\cal A}$ is,
\begin{equation*}
P'((\langle s,q\rangle),(\langle s',q'\rangle)) = \left\{
\begin{array}{ccc}P(s,s'),& $if$\ q'\in\delta(q,L(s'))\\
0,& $otherwise$.
\end{array}
\right.
\end{equation*}}
\end{Definition}

\begin{Remark}\label{sec3:remark7}
{\rm (1) For the definition of $LT$ properties we have assumed that the possibilistic Kripke structure $M$ has no terminal states. It is,
 however, not guaranteed that $M\otimes {\cal A}$ possesses this property, even if $M$ does. This stems from the fact that in $NFA$ ${\cal A}$  there may be a state $q$ that has no direct successor states for some set $u$ of atomic propositions, $i.e$., with $\delta(q,u)=\emptyset$. As we know,  this technical problem can be treated by requiring $\delta(q,u)\neq\emptyset$ for all states $q\in Q$ and input $u$, $i.e$., the finite automaton ${\cal A}$ is \textbf{complete}. Note that imposing the requirement $\delta(q,u)\neq\emptyset$ is not a severe restriction, as any $NFA$ can be easily transformed into an equivalent one that satisfies this property by introducing a state $q_{trap}$ and adding transition $q\overset{u}{\longrightarrow} q_{trap}$ to ${\cal A}$  whenever $\delta(q,u)=\emptyset$ or $q=q_{trap}$.

(2) For each path fragment $\pi=s_0s_1s_2\cdots$ in $M$, since ${\cal A}$ is required complete, there exists at least a run $q_0q_1q_2\cdots$
in ${\cal A}$ for $trace(\pi)=L(s_0)L(s_1)L(s_2)\cdots$such that $q_0\in I$, $q_{i+1}\in\delta(q_i,L(s_i))$ and $\pi^+=\langle s_0,q_1\rangle\langle s_1,q_2\rangle\langle s_2,q_3\rangle\cdots$ is a path fragment in $M\otimes {\cal A}$. The corresponding $\pi^+$ is not unique in general, let us denote the set of all such $\pi^+$ by the symbol $S(\pi)$. The definition of $I'$ and $P'(\langle s,q\rangle,\langle s',q'\rangle)$ guarantees that $Po(\pi)=Po(S(\pi))$. Furthermore, we have the following equality for any $X\subseteq Paths(M)$:
\begin{equation*}
Po^M\{\pi|\pi\in X\}=Po^{M\otimes {\cal A}}(\cup_{x\in X}S(\pi)).
\end{equation*}

        (3) Every path fragment in $M\otimes {\cal A}$ which starts in state $\langle s,\delta(q_0,L(s))\rangle$ arises from the combination of a
 path fragment in $M$ and a corresponding run in ${\cal A}$.}
\end{Remark}

       The following theorem shows that $Po(s\models P_{safe})$ can be derived from the possibility measure of the event $\lozenge B$ in $M\otimes {\cal A}$,
where $B=S\times F$.

   \begin{theorem}
      Let $P_{safe}$ be a regular safety property, ${\cal A}$ be an $NFA$ for the set of good prefixes of $P_{safe}$,
$M$ be a possibilistic Kripke structure, and $s$ be a state in $M$. Then,

$Po^M(s\models P_{safe})=Po^{M\otimes {\cal A}}(\langle s,\delta(q_0,L(s))\rangle\models\lozenge B)$,
  \hfill (8)

\noindent where $B=S\times F$.
\end{theorem}

\begin{proof}
      Let $\Pi$ be the set of paths that start in $s$ and accept $P_{safe}$, $i.e$.,
\begin{equation*}
\Pi=\{\pi\in Paths(s)|Pref(trace(\pi))\cap L({\cal A})\neq\emptyset\}.
\end{equation*}
The set $\Pi^+$ is the set of paths in $M\otimes {\cal A}$ that start in $\langle s,\delta(q_0,L(s))\rangle$ and eventually reach an accept state of ${\cal A}$, $i.e$.,
 \begin{equation*}
 \Pi^+=\{\pi^+\in Paths(\langle s,\delta(q_0,L(s))\rangle)|\pi^+\models\lozenge B\}.
 \end{equation*} By the observation in Remark \ref{sec3:remark7}(2),  for the measurable set $\Pi$ of paths in $M$ and state $s$,
\begin{eqnarray*}
Po_s(\Pi)&=&Po_{\langle s,\delta(q_0,L(s))\rangle}(\{\pi^+|\pi\in\Pi\})\\
&=&Po_{\langle s,\delta(q_0,L(s))\rangle}(\Pi^+)\\
&=&Po^{M\otimes {\cal A}}(\langle s,\delta(q_0,L(s))\rangle\models\lozenge B).
\end{eqnarray*}
\end{proof}

\subsection{Possibility measure of $\omega$-regular property}

     Let us now consider the wider class of $LT$ properties, $i.e$., $\omega$-regular properties. An $LT$ property $P$ is $\omega$-regular whenever $P$
defines an $\omega$-regular language.

     For the $\omega$-regular property $P$,  ${\cal A}$ is assumed to be a B\"{u}chi automaton accepting $P$. We use the symbol $Po(s\models {\cal A})$ to represent $Po(s\models P)$, i.e.,

 $Po(s\models {\cal A})=Po^{M_s}(\{\pi\in Paths(s)|trace(\pi)\in L_\omega({\cal A})=P\})$.\hfill (9)

     It can now be shown, using
similar arguments as for regular safety properties, that the possibility measure of the event $\square\lozenge B$ in the product possibilistic Kripke structure $M\otimes {\cal A}$ coincides with the possibility measure of accepting $P$ by ${\cal A}$. The possibility measure of the event $\square\lozenge B$ can be calculated in polynomial time as shown in Theorem \ref{sec3:theor6}, where $B=S\times F$.

  \begin{theorem}
     Let ${\cal A}$ be an $NBA$ and $M$ a finite possibilistic Kripke structure. Then, for all states $s$ in $M$,

$Po^M(s\models {\cal A})=Po^{M\otimes {\cal A}}\{\langle s,\delta(q_0,L(s))\rangle\models\square\lozenge B\}$,
\hfill (10)

\noindent where $B=S\times F$.
\end{theorem}

\begin{proof}
      The connection between a path $\pi$ in $M$ and the corresponding path $\pi^+$  in $M\otimes {\cal A}$ is as follows,
$trace(\pi)\in L_\omega({\cal A})$ iff $\pi^+\models\square\lozenge B$. By the observation in Remark \ref{sec3:remark7}(2), the possibility measure for path $\pi$ in $M$ with $trace(\pi)\in L_\omega({\cal A})$ agrees with the possibility measure of generating a path $\pi^+$  in $M\otimes {\cal A}$  which arises by the lifting of path  $\pi$ in $M$  where $trace(\pi)\in L_\omega({\cal A})$. The latter agrees with the possibility measure for the paths $\pi^+$  in $M\otimes {\cal A}$ with $\pi^+\models\square\lozenge B$.
\end{proof}

\section{An illustrative example}
We now give an example to illustrate the construction of this paper.

Suppose that there is an animal sicking for a new disease. For the new disease, the doctor has no complete knowledge about it, but he (or she) believes by experience that these drugs such as Ribavirin, Ofloxacin and Thymosin may be useful to the disease.

For simplicity, it is assumed that the doctor considers roughly the animal's condition to be three states, say, ``poor'', ``fair'' and ``excellent''. It is vague when the animal's condition is said to be ``poor'', ``fair'' and ``excellent''. Since the animal's condition can simultaneously belong to ``poor'', ``fair'' and ``excellent'' with respective memberships in the real life situation (\cite{lin02,cao06,liu09}). Therefore, when a possibilistic Kripke structure is used to model the treatment processes of the animal, a fuzzy state is naturally denoted as a three-dimensional vector $[a_1, a_2, a_3]$, which is represented as the possibility distribution of the animal's condition over states ``poor'', ``fair'' and ``excellent''.

Similarly, it is imprecise to say that at what point exactly the animal has changed from one state to another state after a drug treatment (i.e., event), because the drug event occurring may lead a state to multistates with respective membership. Therefore, the treatment process is modeled by a possibilistic Kripke structure, a transition possibility distribution is represented by a $3\times 3$ matrix.

Suppose that the treatment process of the animal is modeled by the following possibilistic Kripke structure $M=(S,P,I,AP,L)$, where
$S=AP=\{poor, fair$, $excellent\}$,

$P=\left(\begin{array}{cccc}
0.5&1&0.5\\
0.2&0.5&1\\
0.2&1&1
\end{array}
\right)$,
$I=\left(\begin{array}{cccc}
1\\
0\\
0
\end{array}
\right)$,

\noindent and $L(s)=\{s\}$ for any $s\in S$.

The structure $M$ is presented in Fig. 4, and the corresponding $M^+$ is presented in Fig. 5, where we use the symbols $p, f, e$ to represent the states or the atomic propositions ``poor'', ``fair'' and ``excellent'' respectively.

\begin{figure}[ht]
\begin{center}
\includegraphics[scale=0.5]{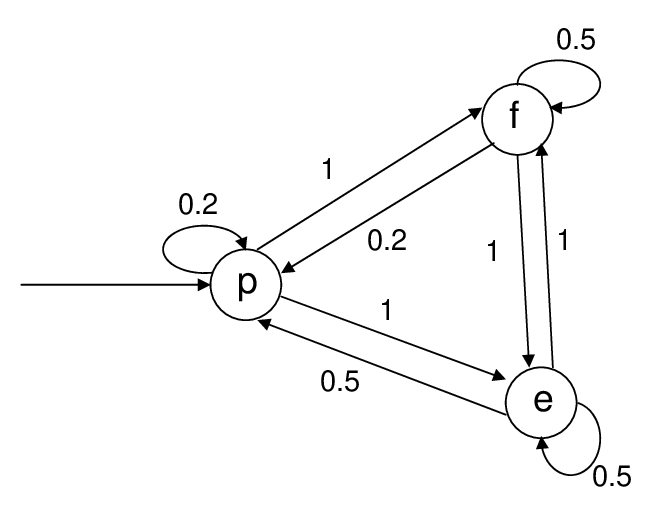}
\center{Fig.4.}The possibilistic Kripke structure $M$ for the treatment process of the animal.
\vspace{-0.3cm}
\end{center}
\end{figure}

\begin{figure}[ht]
\begin{center}
\includegraphics[scale=0.5]{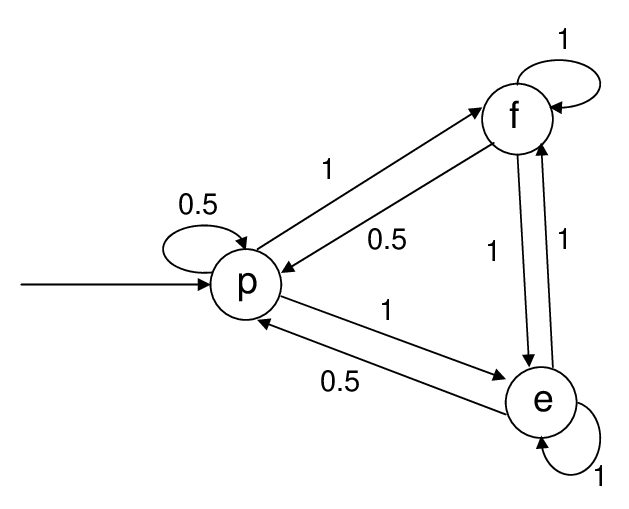}
\center{Fig.5.}The corresponding $M^+$ of $M$ in Fig.4.
\vspace{-0.3cm}
\end{center}
\end{figure}

Let us do some model checking using the above structure $M$.

First, let us calculate $Po(\lozenge\{excellent\})$. Using Eq. (1), we have,

$Po(\lozenge\{excelent\})=\vee_{s\in S}I(s)\wedge P^+(s,excellent)=1$.

Using Eq. (7), we have,

$Po(poor\models\square\lozenge\{excelent\})=P^+(poor,excelent)\wedge P^+(excelent,excelent)=1\wedge 1=1$.

Consider a regular safety property $P_{safe}$ with good prefixes accepted by an NFA ${\cal A}$ as shown in Fig. 6, here $GPref(P_{safe})=(2^{AP})^*\{excelent\}$, which represents the property ``the drug will eventually be useful for the disease''.

\begin{figure}[ht]
\begin{center}
\includegraphics[scale=0.5]{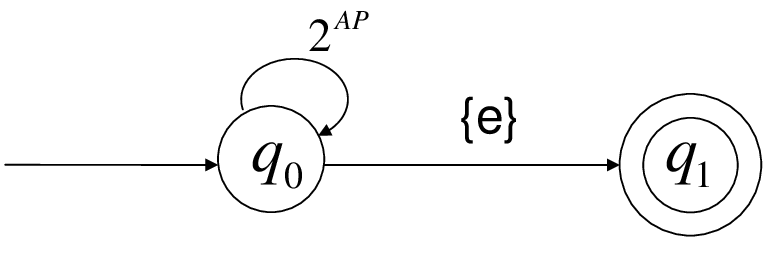}
\center{Fig.6.}The NFA ${\cal A}$ accepts $GPref(P_{safe})=(2^{AP})^*\{excelent\}$.
\vspace{-0.3cm}
\end{center}
\end{figure}

Let us calculate $Po(s\models P_{safe})$, where $s=poor$. The product possibilistic Kripke structure $M\otimes {\cal A}$ is shown in Fig. 7.

\begin{figure}[ht]
\begin{center}
\includegraphics[scale=0.5]{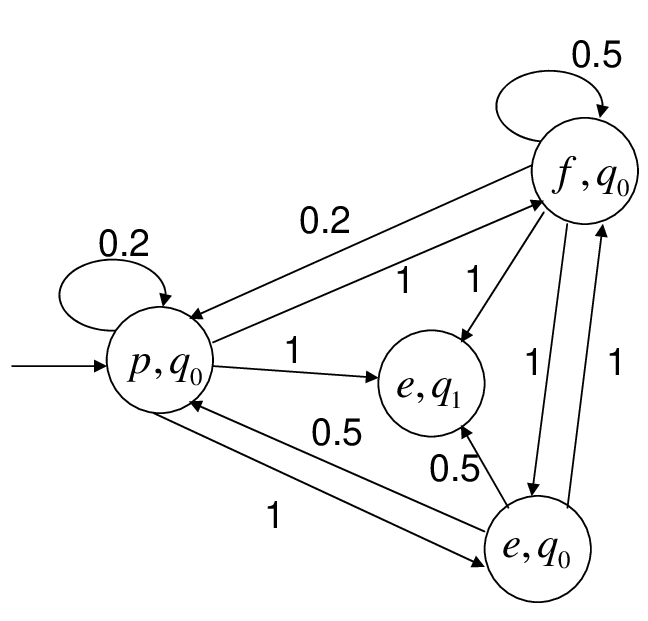}
\center{Fig.7.}The product possibilistic Kripke structure $M\otimes {\cal A}$.
\vspace{-0.3cm}
\end{center}
\end{figure}

Using Eq. (8), we have

$Po(poor\models P_{safe})=Po^{M\otimes {\cal A}}(\langle poor,\delta(q_0, poor)\rangle\models\lozenge B)=Po^{M\otimes {\cal A}}(\langle poor,q_0\rangle\models\lozenge B)=1$.

It shows that the possibility of the event ``the drug will eventually be useful for the disease'' is very high, it is $1$.

Next, let us consider an $\omega$-regular property $L=\{poor\}\{poor\}^{\omega}$ which can be accepted by an NBA ${\cal B}$ as shown in Fig. 8. Here $L$ represents a property ``the drug is useless for the disease''. The product possibilistic Kripke structure $M\otimes {\cal A}$ is shown in Fig. 9.

\begin{figure}[ht]
\begin{center}
\includegraphics[scale=0.5]{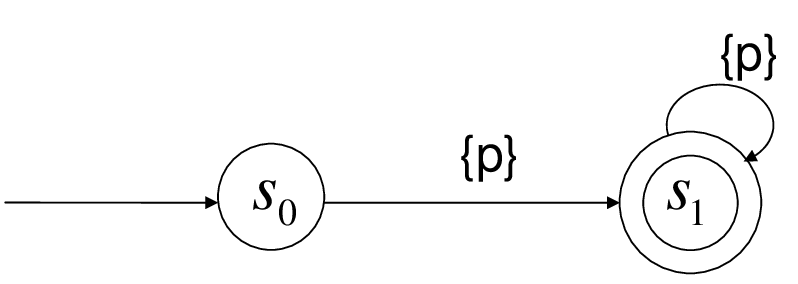}
\center{Fig.8.}The NBA ${\cal A}$ accepts $L=\{poor\}\{poor\}^{\omega}$.
\vspace{-0.3cm}
\end{center}
\end{figure}

\begin{figure}[ht]
\begin{center}
\includegraphics[scale=0.5]{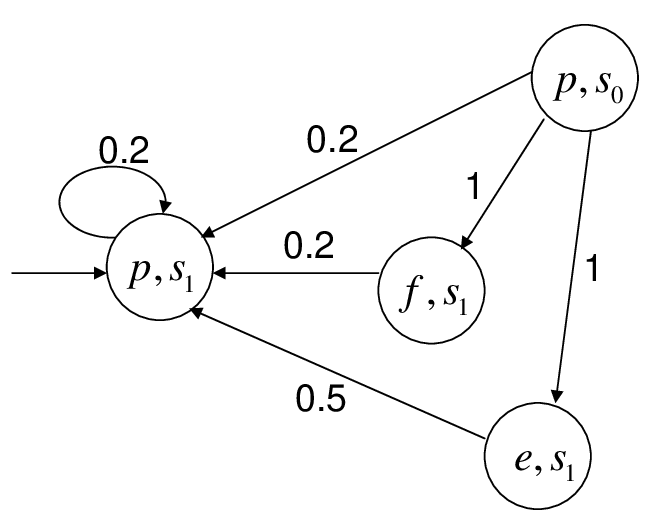}
\center{Fig.9.}The product possibilistic Kripke structure $M\otimes {\cal B}$.
\vspace{-0.3cm}
\end{center}
\end{figure}

Using Eq. (10), we have,

$Po^M(poor\models {\cal A})=Po^{M\otimes {\cal A}}\{\langle poor,s_1\rangle\models\square\lozenge B\}=0.2$.

It shows that the possibility of the event ``the drug is useless for the disease'' is very low, it is $0.2$.

\section{Conclusions}\label{sec6:con}
	$LTL$ model checking based on possibility measure is a fuzzy measure extension of classical  model checking. Both the possibilistic and probabilistic model-checking solve certain uncertainty of error or other stochastic behavior occurring in various real world applications. In this paper, we studied several important possibility measures of $LT$ properties and $LTL$ formulae corresponding to them. Concretely, we introduced the notions of $LT$ properties; several particular $LT$ properties such as reachability and repeatedly reachability were introduced. More generally, $LT$ properties such as regular safety properties, $\omega$-regular properties using automata theory were studied. In fact, we introduced the product possibilistic Kripke structure of a possibilistic Kripke structure and a finite automaton. In which, the computation of possibility measure of possibilistic Kripke structure meeting $LT$ property can be translated into reachability possibility or repeated reachability possibility of the product possibilistic Kripke structure. With these notions, we gave the quantitative verification methods of regular safety properties and $\omega$-regular properties.

This is an initial work on the model checking using possibility measure. There are many things to be done along this direction.

\begin{itemize}
\item We use max-min composition of fuzzy relations in this paper. There are other forms of composition of fuzzy relations, such as max-product composition. They may be more appropriate in some real world applications of fuzzy sets than max-min composition. Then the related work using other composition instead of max-min composition can be done in the future.

\item We use the normal possibility distribution in this paper (see conditions (2) and (3) in the definition of possibilistic Kripke structure). These restrictions are too strict for some applications of the  method proposed in this paper. It is natural to relax these restrictions and to use generalized possibility measure in the model-checking.

\item The properties considered in this paper are classical, we can further consider the properties with fuzzy uncertainties. In this case, we can use fuzzy automata (see for example \cite{li052}) instead of classical automata to describe the related properties of systems.

\item As we know, there has been many work on model checking of $LT$ properties in multi-valued systems, see for example \cite{Li11} and references therein. In our future work, we shall give some comparisons of our method with the methods in multi-valued model checking \cite{Li11}.

\item Another direction is to extend the method used in this paper to the $CTL$ model checking in possibilistic Kripke structure.

\end{itemize}

Of course, the most important thing is to give some case studies of the methods proposed in this paper.


%



 \section*{Acknowledgments}

The authors would like to thank the anonymous referees for helping them refine the ideas
presented in this paper and improve the clarity of the presentation. The authors would also like to express their special thanks to Dr. Sanjiang Li at University of Technology Sydney for detailed
suggestions that improved the paper's quality.

\begin{IEEEbiography}[{\includegraphics[width=1in,height=1.25in,clip,keepaspectratio]{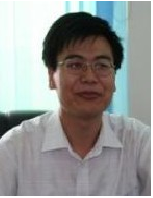}}]{Yongming Li}
received the Ph.D. degree in mathematics
from Sichuan University, Chengdu, China, in
1996.
He is currently a Professor of mathematics and
computer science at Shaanxi Normal University,
Xi'an, China. 

His research interests include computation
theory, fuzzy control theory, fuzzy automata
theory, spatial reasoning, quantum logic and quantum computation,
and topology over lattices.
\end{IEEEbiography}

\begin{IEEEbiography}[{\includegraphics[width=1in,height=1.25in,clip,keepaspectratio]{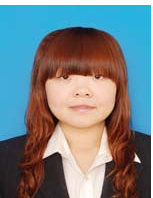}}]{Lijun Li}
received the master degree in computer science
from Shaanxi Normal University, Xi'an, China, in
2012.
She is currently a teacher of
computer science at Taiyuan No. 66 Middle School, Taiyuan, China. 

Her research interests include computation
theory and  model checking.
\end{IEEEbiography}





\end{document}